%% file: lrgn_two.tex
\documentstyle[prd,aps,epsf,eqsecnum,twocolumn,subfigure]{revtex}
\input{title2.tex}

\newcommand{\C}{{\cal C}}
\newcommand{\real}{{\cal R}e}
\newcommand{\imag}{{\cal I}m}
\begin{document}
%
%%%%%%%%%%%%%%%%%%%%%%%%%%%%%%%%%%%%%%%%%%%%%%%%%%%%%%%%%%%%%%%%%%%%%%
%
\title{Order $1/N$ corrections to the time-dependent Hartree
       approximation for a system of $N+1$ oscillators}
\author{Bogdan Mihaila\thanks{electronic mail:Bogdan.Mihaila@unh.edu} and
        John~F.~Dawson\thanks{electronic mail:John.Dawson@unh.edu}}
\address{Department of Physics, University of New Hampshire,
Durham, NH 03824}
\author{Fred Cooper\thanks{electronic mail:Cooper@pion.lanl.gov}}
\address{Theoretical Division, MS B285, Los Alamos National
Laboratory, Los Alamos, NM 87545}
\date{\today}
\preprint{LA-UR-97-1673}
\abstract{
We solve numerically to order $1/N$ the time evolution of a quantum
dynamical system of $N$ oscillators of mass $m$ coupled quadratically
to a massless dynamic variable.  We use Schwinger's closed time path
(CTP) formalism to derive the equations.  We compare two methods which
differ by terms of order $1/N^2$.  The first method is a direct
perturbation theory in $1/N$ using the path integral.  The second
solves exactly the theory defined by the effective action to order
$1/N$. We compare the results of both methods as a function of $N$.
At $N=1$, where we expect the expansion to be quite innacurate, we
compare our results to an exact numerical solution of the
Schr\"odinger equation.  In this case we find that when the two
methods disagree they also diverge from the exact answer. We also find
at $N=1$ that the $1/N$ corrected evolutions track the exact answer
for the expectation values much longer than the mean field ($N=
\infty$) result.
}
\pacs{11.15.Pg, 11.30.Qc, 25.75.-q, 3.65.-w}
\maketitle2 
%\narrowtext
%
%%%%%%%%%%%%%%%%%%%%%%%%%%%%%%%%%%%%%%%%%%%%%%%%%%%%%%%%%%%%%%%%%%%
%
\section{Introduction}

The large $N$ approximation has a long history in both statistical
mechanics and quantum field theory, mostly in determining the phase
structure of various theories~\cite{ref:largeN}. It is only recently
that this approximation has been used to study the dynamical evolution
of various systems, and at present only the lowest order in the large
$N$ expansion has been considered.  In leading order, the large $N$
expansion is equivalent to using a gaussian density matrix, and
therefore two particle scattering effects are included only
indirectly.  The leading order in the large $N$ approximation is
closely related to a time dependent Hartree approximation.  The exact
connection between these methods is discussed in detail in
ref.~\cite{ref:CDHKMS}.  Although interesting results in lowest order
have been obtained for pair production from strong
fields~\cite{ref:KESCM,ref:CEKMS} as well as the evolution of a chiral
phase transition~\cite{ref:chiral}, the important effects of the direct
two particle elastic scattering, which determines the thermalization
time scale of the plasma, is not included in the mean field
approximation.  In order to compare time scales for rethermalization
with, say, the plasma oscillation frequency, as well as the expansion
time of an evolving plasma produced in a heavy-ion collision, one
needs to go to next order in the $1/N$ expansion.  The property of the
$1/N$ expansion relevant here is that connected $2n$-point Green's
functions first appear with $G_{2 n}$ of order $1/N^{n-1}$.  Thus
direct scattering of two particles first occurs at order $1/N$.  At
lowest order in $1/N$, the equations one has to solve are differential
equations.  At next order one gets integro-differential equations
which depend on the time history of the system.  This requires new
numerical methods to ensure that the update conserves energy.

There are two different ways one can determine the $1/N$ corrections.
The first method is to iterate the solution of the lowest order
calculation in a standard perturbative fashion.  However, one might
hope that it might be more accurate to first Legendre transform the
action obtained to order $1/N$ and to obtain a new effective action,
which differs by terms of order $1/N^2$ from the first method.  In the
second method, one evolves directly the equations of motion obtained
from the effective action.  By having these two different methods, one
has an upper bound on the accuracy of the $1/N$ expansion. When the
two methods diverge, this is a signal that $1/N^2$ corrections are
important.  At $N=1$ this divergence is very close to the place where
these two methods diverge from the exact answer. We find that the
method based on the effective action is actually less stable, since
solutions become unbounded earlier.  However this occurs much after
the method is unreliable.

In quantum field theory it is not possible to compare the $1/N$
expansion with an exact calculation of a dynamical evolution because
of the large number of degrees of freedom.  Thus we thought it
appropriate to study a simple quantum mechanical example which we have
studied before in the lowest order mean field
approximation~\cite{ref:CDHKMS,ref:CDMS}.  The advantage of this simple
model is that, at least for $N=1$, comparisons can be made with a
direct numerical simulation of the exact problem~\cite{ref:CDHR}.
Unfortunately, even for the quantum mechanics case, going beyond $N=1$
is not numerically feasible for the exact problem.  So we are testing
our large $N$ expansion in a regime where it is not expected to work
very well.  However, in spite of this short coming, we find that by
adding the $1/N$ correction terms, our approximation tracks the exact
answer for expectation values a factor of two longer than the lowest
order approximation --- which is encouraging.

As far as we know, this is the first attempt to use Schwinger's CTP
formalism in a calculation which is not a perturbation expansion in
the coupling constant.  The methods of solving the resulting
Volterra-like equations, which we present here, can be generalized to
the field theory case. Therefore this toy model, which can be compared
with a direct solution of the Schr\"odinger equation, is an ideal
problem to test the accuracy of the numerical methods needed for 
field theory calculations.

%
%%%%%%%%%%%%%%%%%%%%%%%%%%%%%%%%%%%%%%%%%%%%%%%%%%%%%%%%%%%%%%%%%%%
%
\section{The generating functional}
\label{sec:II}

We consider a system of $N$ oscillators of mass $m$ coupled
quadratically to a massless oscillator with coupling $e$.  This
quantum mechanical system is a model of a single momentum mode of
scalar quantum electrodynamics~\cite{ref:CHKMPA}.  The Lagrangian for
this system is given by:
\begin{eqnarray}
   L 
   & = & 
   \frac{1}{2} \dot{A}^2 + J A
   \nonumber \\ &&
   + \sum_{a=1}^{N} \left\{   
       \frac{1}{2} \, \dot{\phi}_a^2 - \frac{1}{2} ( m^2 + e^2 A^2)
                   \, \phi_a^2 + j_a \phi_a \right\}
   \>.
\label{eq:Lag}
\end{eqnarray}
We introduce the scaled variables, defined as
\begin{equation}
   \begin{array}{rclrcl}
      A      & \rightarrow & A/\sqrt{N} \>, \qquad &
      \phi_a & \rightarrow & \phi_a/\sqrt{N}, \quad a = 1,N \\
      J      & \rightarrow & J/\sqrt{N} \>, \qquad &
      j_a    & \rightarrow & j_a/\sqrt{N},       \quad a = 1,N \\
      e      & \rightarrow & e \sqrt{N} \>. \qquad &
      L      & \rightarrow & L / N
   \end{array} \\
\label{eq:scaling}
\end{equation}
From now on, we use scaled variables.  We wish to consider the time
evolution of expectation values of observables for an initial value
problem, $0 < t < \infty$.  The way to formulate an initial value
problem in quantum mechanics, using a generating functional, was done
more than thirty years ago by Schwinger, Bakshi and Mahanthappa, and
later by Keldysh~\cite{ref:SBMK}.  This formalism, which is in the
Heisenberg picture, is related to the fact that in the Schr\"odinger
picture the evolution of the density matrix in quantum mechanics,
\begin{equation}
   \hat{\rho}(t) \ = \
      e^{-i\hat{H}t} \, \hat{\rho}(0) \, e^{ i\hat{H}t}
   \>,
\end{equation}
requires both a forward evolution from zero to $t$ and a backward one
from $t$ to zero.  The average value of observables are given by
traces over states of the system:
\begin{displaymath}
   \langle \, \hat{O}(t) \, \rangle
      =  {\rm Tr} \{ \, \hat{\rho}(t) \, \hat{O} \, \}
      =  {\rm Tr} \{ \, \hat{\rho}(0) \, \hat{O}(t) \, \}
   \>.
\end{displaymath}
in the Schr\"odinger and Heisenberg picture, respectively.

\begin{figure}
   \epsfxsize = 3.0in
   \centerline{\epsfbox{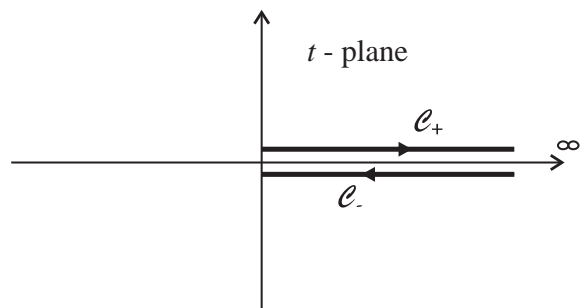}}
   \caption{Complex time contour $\C$ for the closed time path
            integrals.}
   \label{fig:CTP}
\end{figure}

As explained in the Appendix of ref.~\cite{ref:CDHKMS}, this
necessitates both positive and negative time ordered operators in the
evolution of the observable operators and the introduction of two
currents into the path integral for the generating functional.  A
concise way of writing the needed functions is to define the time
integrals along a path in the complex time plane.  This closed time
path (CTP) is shown in Fig.~\ref{fig:CTP}.  The CTP integration
contour is given by:
\begin{equation}
   \int_{\C} F(t) \, {\rm d}t \ = \
      \int_{0:\C_{+}}^{\infty} F_{+}(t) \, {\rm d}t -
      \int_{0:\C_{-}}^{\infty} F_{-}(t) \, {\rm d}t  \>.
\end{equation}
Using the CTP contour, the generating functional for the causal
Green's functions for the theory described by the Lagrangian
(\ref{eq:Lag}) is given by the path integral:
\begin{eqnarray}
   Z[J,j]
   & = & \int\! {\rm d} [A]
         \int\! {\rm d} [\phi_a]
      \, e^{ i  N \, S[A,\phi;J,j] }
   \nonumber \\
   S[A,\phi; J,j] & = & \int_{\C} {\rm d}t \, L
   \>,
\label{eq:ZSL}
\end{eqnarray}
The full closed time path Green's function for the two point functions
is:
\begin{displaymath}
   G_{ab}(t,t') \ = \
      G_{ab}^{>}(t,t') \, \Theta_{\C}(t,t') +
      G_{ab}^{<}(t,t') \, \Theta_{\C}(t',t)  \>,
\end{displaymath}
in terms of the Wightman functions,
\begin{eqnarray*}
   G^{>}_{ab}(t,t') & = & 
      i \{ \langle \phi_a(t)\phi _b(t') \rangle - 
           \langle \phi_a(t) \rangle \langle \phi_b(t') \rangle 
        \}  
   \\
   G^{<}_{ab}(t,t') & = & 
      i \{ \langle \phi_b(t') \phi_a(t) \rangle - 
           \langle \phi_b(t') \rangle \langle \phi_a(t) \rangle
        \} \>,
\end{eqnarray*}
where $\langle \phi_a(t) \phi_b(t') \rangle \equiv {\rm Tr} \{ \rho(0) \,
\phi_a(t) \phi_b(t')\}$, and where the CTP step function 
$\Theta_{\C}(t,t')$ is defined by:
\begin{equation}
   \Theta_{\C}(t,t') =
   \left \{
      \begin{array}{ll}
         \Theta(t,t') &
            \mbox{for $t$ on $\C_{+}$ and $t'$ on $\C_{+}$} \\
         0            &
            \mbox{for $t$ on $\C_{+}$ and $t'$ on $\C_{-}$} \\
         1            &
            \mbox{for $t$ on $\C_{-}$ and $t'$ on $\C_{+}$} \\
         \Theta(t',t) &
            \mbox{for $t$ on $\C_{-}$ and $t'$ on $\C_{-}$}
      \end{array}
   \right .
\end{equation}
This is equivalent to a $2 \times 2$ matrix Green's function on the
vector space $\{ +,- \}$, often found in the literature.

The large N expansion is obtained by performing the Gaussian integral
over the $\phi_a$ variables to obtain an effective action, and
evaluating the remaining integral over $A$ by the method of steepest
descent.  This gives:
\begin{eqnarray}
   Z[J,j] & = & \int d[A] e^{i N  S_{{\rm eff}}[A ; J,j] } 
   \equiv  e^{i N \, W[J,j]}  \nonumber \\
   W[J,j] & = &
    S_{{\rm eff}}[A_0 ; J,j]
     +  \frac{i}{2N} \int_{\C} {\rm d}t   \ln [ D^{-1}(t,t) ]
   \nonumber \\ &&
     + \cdots \>.
\label{eq:ZW}
\end{eqnarray}
where $S_{{\rm eff}}[A_0 ; J,j]$ is given by
\begin{eqnarray}
  &&
  S_{{\rm eff}}[A_0 ; J,j] \ = \ 
  \nonumber \\ &&
  \int_{\C} {\rm d}t \,
     \biggl \{
       - \frac{1}{2}
         A_0 \frac{ {\rm d}^2 }{ {\rm d} t^2 } A_0
       + J A_0
       + \frac{i}{2N} \sum_{a=1}^{N} \ln [ G^{-1}_{0\,aa}(t,t) ]
     \biggr \}
  \nonumber \\ &&
  + \frac{1}{2} \int_{\C} {\rm d}t \, \int_{\C} {\rm d}t' \,
    \sum_{a,b=1}^{N} \, j_a(t) \, G_{0\,ab}(t,t') \, j_b(t')
  \>.
\label{eq:Szero}
\end{eqnarray}
The stationary point $A_0(t)$ is determined by
\begin{equation}
 \left\{ \frac{ {\rm d}^2 }{ {\rm d}t^2 }
   \ + \ e^2 \sum_{a=1}^{N} \left[
             \phi_{0\,a}^2(t) + \frac{1}{iN} G_{0\,aa}(t,t)
             \right] \right\} A_0(t)
    =  J(t).
 \label{eq:As}
\end{equation}
Here $\phi_0$ is defined as:
\begin{equation}
  \left  \{
      \frac{ {\rm d}^2 }{ {\rm d} t^2 } + ( m^2 + e^2 A_0^2)
   \right \} \, \phi_{0\,a}(t)
    =  j_a(t)  \>,
\label{eq:phi0}
\end{equation}
and $G_0$ is given by the solution to
\begin{equation}
 \left  \{
      \frac{{\rm d}^2}{{\rm d} t^2} + ( m^2 + e^2 A_0^2)
   \right \}
      G_{0\,ab}(t,t')
  =
   \delta_{ab} \ \delta_{\C}(t,t')  \>.
\label{eq:G0}
\end{equation}
$\phi_{0\,a}(t)$ and $A_0(t)$ are to be regarded as functionals of the
sources, $J(t)$ and $j(t)$.  We have defined $\delta_{\C}(t,t') = {\rm
d} \Theta_{\C}(t,t') / {\rm d} t$.

The inverse  propagator $D^{-1}(t,t')$ is
\begin{eqnarray}
   D^{-1}(t,t') & = & 
   - \left[\frac{ \delta^2 S_{{\rm eff}}[A ; J,j]}
     {\delta A(t) \delta A(t')} \right]_{A_0}
   \nonumber \\ & = &
   D_0^{-1}(t,t') + \Pi_0(t,t')
   \>,
\label{eq:Dinv}
\end{eqnarray}
where
\begin{eqnarray}
   &&
   D_0^{-1}(t,t') \ = \ 
   \nonumber \\ && 
       \left\{ \frac{ {\rm d}^2 }{ {\rm d}t^2 }
               +  e^2 \sum_{a=1}^{N}
                        \left  [
              \phi_{a}^2(t) + \frac{1}{iN} G_{0\,aa}(t,t)
                         \right ]
       \right\} \, \delta_{\C} (t,t') 
\label{eq:D0} 
\end{eqnarray}
and
\begin{eqnarray}
   &&
   \Pi_0(t,t') = 2 e^4 \, A_0(t) A_0(t')
       \sum_{a,b=1}^{N}
       \biggl \{ \frac{i}{N} \,
          G_{0\,ab}(t,t') G_{0\,ba}(t',t)
   \nonumber \\ && \quad \quad \quad \quad \quad \quad \quad 
          - 2 \,
          \phi_{0\,a}(t) \, G_{0\,ab}(t,t') \, \phi_{0\,b}(t')
       \biggr \} 
   \>.
\label{eq:Pi0}
\end{eqnarray}
We solve (\ref{eq:G0}) by introducing a complete set of solutions
$f(t)$ to the homogeneous equation, satisfying the Wronskian condition
\begin{equation}
   f^{\ast}_{a}(t) \dot{f}_{a}(t) -
   \dot{f}^{\ast}_{a}(t) f_{a}(t)
      \ = \ - i.
\end{equation}
The causal Green's functions $G_0$ can then be written:
\begin{eqnarray}
   &&
   G_{0\,ab}(t,t') \ = \
   \nonumber \\ && 
      i \delta_{ab}
      \left \{
         f_{a}(t) f_{a}^{\ast}(t') \, \Theta_{\C}(t,t') +
         f_{a}(t') f_{a}^{\ast}(t) \, \Theta_{\C}(t',t)
      \right \}
   \>.
\end{eqnarray}

%
%%%%%%%%%%%%%%%%%%%%%%%%%%%%%%%%%%%%%%%%%%%%%%%%%%%%%%%%%%%%%%%%%%%
%
\section{Time Evolution Equations}

In the Schr\"odinger picture the Schr\"odinger equation governs the
time evolution of the $N+1$ oscillators :
\begin{eqnarray}
   i \frac{\partial \Psi(\phi,A,t)}{\partial t} 
   & = & 
   \biggl  \{ 
      \sum_a \, \frac{1}{2} \, 
          \biggl [
             - \, \frac{\partial^2}{\partial \phi_a^2}
             \, + \, (m^2 + e^2 A^2) \phi_a^2
          \biggr ]
   \nonumber \\ && \quad \quad \quad \quad 
      - \frac{1}{2} \frac{\partial^2}{\partial A^2}
   \biggr \} \, 
   \Psi (\phi,A,t)
   \>.
\end{eqnarray}
It is the solution of this equation that we will compare with our
large $N$ equations for the expectation values.  Since at leading
order in large $N$ an initial Gaussian wave packet stays Gaussian
under time evolution, we will start our problem with Gaussian initial
data.  Also we need to relate the parameters of the wave function at
time zero to the values of one and two point functions and their time
derivatives, since the Green's functions obey second order
differential equations.  At $t=0$ we choose our wave function to be a
product of Gaussians:
\begin{equation}
   \Psi(\phi,A,0) = \Psi_\phi(0) \, \Psi_A(0) \>,
   \label{eq:wave}
\end{equation}
where
 \[
   \Psi_\phi(0)= [2 \pi G(0)]^{-1/4} \, \exp
      \left  [ 
         -x^2 \left  ( G^{-1}(0)/4  - i \Pi_G(0)
              \right )
      \right ]
\]
and
\begin{eqnarray*}
   \Psi_A(0) & = & [2 \pi D(0)]^{-1/4} \, 
   \nonumber \\ &&
   \exp
      \biggl [ 
         - \, ( A - \sqrt{N}\tilde{A}_0 )^2 \,
         ( D^{-1}(0)/4 - i \Pi_D(0) ) 
   \nonumber \\ && \quad \quad 
         + \,
         i p_A(0) ( A - \sqrt{N}\tilde{A}_0 )
      \biggr ]
\end{eqnarray*}
The variables have the following meaning:
\begin{eqnarray*}  
   \langle \phi_a(0) \phi_b(0) \rangle & = & \delta_{ab} G(0)  \\
    \Pi_G(0)) & = & \frac{\dot{G}(0)}{4 G(0)}
\end{eqnarray*}
We have chosen $\langle \phi_a \rangle = \langle {\dot \phi}_a \rangle
= 0$.
\begin{eqnarray*}
   D(0)+ N \tilde{A}_0^2 & = & \langle A^2 \rangle_{t=0} \\
   \Pi_D(0) & = & \frac{\dot{D}(0)}{4 D(0)}    \\
   p_A(0) & = & \langle -i \partial/\partial A \rangle_{t=0} 
            = \langle {\dot A}(0) \rangle 
\end{eqnarray*}
we will evolve these equations (for $N=1$) using a symplectic
integrator as described in reference~\cite{ref:CDHR} to compare with
the results of the large $N$ expansion.  The parameter $G(t)$ in the
Schr\"odinger wave function which is the real part of the width of the
wave function is related to $G_0(t,t')$ of the Heisenberg approach via
$G(t) = G_0(t,t)/i$.

%
%%%%%%%%%%%%%%%%%%%%%%%%%%%%%%%%%%%%%%%%%%%%%%%%%%%%%%%%%%%%%%%%%%%
% 
\subsection{Leading order in $ 1/N$}

Let us now consider the equations and the initial conditions for the
large N expansion.  For simplicity and also because we are considering
these equations as the single mode approximation to scalar
electrodynamics we consider the case where $\phi_{0 \, a} = 0$ for all
$t$. In terms of the mode functions $f_a$ discussed above we then
have:
\begin{equation}
   G_{0\,ab}(t,t)/i \ =
     \delta_{ab}  f_{a}(t) f_{a}^{\ast}(t)
\end{equation}
The coupled equations which need to be solved are:
\begin{eqnarray}
   \left\{ \frac{ {\rm d}^2 }{ {\rm d}t^2 }
     \ + \ \frac{e^2}{N} \sum_{a=1}^{N} | f_{a}(t) |^2
   \right\} A_0(t)
   & = & 0
\label{eq:A0} \\
   \left\{
      \frac{ {\rm d}^2 }{ {\rm d} t^2 } \ + \ [ m^2 + e^2 A_0^2(t) ]
   \right\}  f_{a}(t) & = & 0  \>,
\label{eq:f}
\end{eqnarray}
The four initial conditions that have to be specified in lowest order
are
\begin{displaymath} 
   A(0); \quad \dot{A}(0) ; \quad G(0), \quad \dot{G}(0)  \>. 
\end{displaymath}
Here we have assumed that the $A$ field can be treated classically,
which is the limit where the width of the fluctuations in $A$, namely
$D$, can be ignored.  In general we also have to specify $D(0)$ and
$\dot{D}(0)$.  In terms of the $f_a$ this translates into the initial
conditions
\begin{eqnarray*}
   f_{a}(0) & = & \sqrt{G(0)} \\
   \dot{f}_{a}(0) & = & 
      \left  [ - \frac{i}{2 G(0)} + \frac{ \dot{G}(0) }{ 2G(0) }
      \right ] \, f_{a}(0) \>.
\end{eqnarray*}
for all $a$.

%
%%%%%%%%%%%%%%%%%%%%%%%%%%%%%%%%%%%%%%%%%%%%%%%%%%%%%%%%%%%%%%%%%%%
% 
\subsection{Order $1/N$ corrections}

The $1/N$ corrections to the generating functional $Z$ require the
evaluation of the second term in (\ref{eq:ZW}).  We invert
(\ref{eq:Dinv}) and find:
\begin{eqnarray}
   &&
   D(t,t') \ = \ D_0(t,t') 
   \nonumber \\ && \quad \quad \quad
   - \ \int_{\C} {\rm d}t_1 \, \int_{\C} {\rm d}t_2 \,
       D_0(t,t_1) \, \Pi_0(t_1,t_2) D(t_2,t')
   \>.
\label{eq:Dfcn}
\end{eqnarray}
Since we have chosen $\phi_{0\,a}(t) = 0$, we can solve $D_0(t,t')$ as
follows.  Find the two linearly independent solutions $g$ and
$g^{\ast}$ to the homogeneous equation:
\begin{equation} 
   \left  \{
      \frac{ {\rm d}^2 }{ {\rm d}t^2 } \ + \ 
      \frac{e^2}{N} \sum_{a=1}^{N} | f_{a}(t) |^2 
   \right \} \, g(t) \ = \ 0 \>.
\label{eq:g}
\end{equation}
satisfying the Wronskian condition:
\begin{equation}
   g^{\ast}(t) \, \dot{g}(t) - \dot{g}^{\ast}(t) \, g(t)
      \ = \ - i  \>,
\end{equation}
In terms of these solutions we have:
\begin{equation}
   D_0(t,t')
   \ = \ i \{
      g(t) \, g^{\ast}(t') \Theta_{\C}(t,t') \ + \
      g(t') \, g^{\ast}(t) \Theta_{\C}(t',t)  \}
\end{equation}
The initial width of the wave function is then given by
\begin{equation}
   D(0) = D_0(0,0) /i = | g(0) |^2
\end{equation}
Thus we can relate the initial conditions on $g$ to the initial
conditions $D(0)$ and $\dot{D} (0)$ of the wave function as follows:
\begin{eqnarray*}
   g(0) & = & \sqrt{D(0)} \\
   \dot{g} (0) & = & 
      \left  [ 
         - \frac{i}{2 D(0)} + \frac{{\dot D}(0)}{2D(0)} 
      \right ] \, g(0).
\end{eqnarray*}
$\Pi_0(t,t')$ is given by:
\begin{eqnarray}
   && \Pi_0(t,t') \ = \
       - 2 i e^4 \,
       A_0(t)  A_0(t')
   \nonumber \\ &&
   \biggl \{
      \frac{1}{N} \sum_{a=1}^{N} 
      \Bigl [ f_a^2(t) \, {f_a^{\ast}}^2(t') \Theta_{\C}(t, t') 
          +  f_a^2(t') \, {f_a^{\ast}}^2(t) \Theta_{\C}(t', t) 
      \Bigr ]  
   \biggr \}
   \>.
   \nonumber \\
\end{eqnarray}
Our numerical strategy for solving  eq.~(\ref{eq:Dfcn}) is discussed in
appendix \ref{sec:A}.

There are two ways to calculate the order $1/N$ corrections to the
time evolution problem.  The first way is by a straight forward
perturbation expansion of $W[J,j]$ in powers of $1/N$.  If we consider
the average value of $A(t)$, we have:
\begin{equation}
   A(t)
   \ = \ \frac{1}{i N \, Z[j,j]}
         \left[ \frac{ \delta Z[J,j] }{ \delta J(t) } \right]_{j}
         \ = \ A_0(t) + \frac{1}{N} \, A_1(t)   \>,
\end{equation}
where
\[ 
   A_1(t) =  \left[ \frac{ \delta  }{ \delta J(t) } \right]  
   \frac{i}{2} \int_{\C}
   {\rm d}t' \,  \ln [ D^{-1}(t',t') ]
\]
Computing the derivatives, we obtain:
\begin{equation}
   A_1(t) \  = \ A_1^{(a)}(t) + A_1^{(b)}(t) + A_1^{(c)}(t)
      \>.
\label{eq:A1}
\end{equation}
where
\begin{eqnarray*}
   A_1^{(a)}(t)
   & = & - e^4 \
         \int_{\C} {\rm d}t' \, D(t,t') \, A_0(t') \Sigma_1(t')
   \\
   A_1^{(b)}(t)
   & = & - 2 e^4 \
         \int_{\C} {\rm d}t' \, D(t,t') \, \Omega(t')
   \\
   A_1^{(c)}(t)
   & = & 4 e^6 \
         \int_{\C} {\rm d}t' \, D(t,t') \, A_0(t') \Sigma_2(t')
   \>,
\end{eqnarray*}
where we have introduced
\begin{eqnarray*}
   \Sigma_1(t)
   & = & \frac{1}{N} \sum_{a,b=1}^{N}
      \int_{\C} {\rm d}t' \,
      G_{0\,ab}(t,t') \, D(t',t') \, G_{0\,ba}(t',t)
   \\
   \Sigma_2(t)
   & = & \frac{1}{N} \sum_{a,b,c=1}^{N}
   \int_{\C} {\rm d}t' \, 
       G_{0\,ac}(t,t') A_0(t') \, 
   \nonumber \\ &&
   \int_{\C} {\rm d}t'' \,
       G_{0\,cb}(t',t'') \, D(t',t'') \, 
       A_0(t'') \,
       G_{0\,ba}(t'',t)
   \\
   \Omega(t)
   & = & \frac{1}{N} \sum_{a,b=1}^{N}
   \int_{\C} {\rm d}t' \, G_{0\,ab}(t,t') D(t,t') 
   \nonumber \\ && \qquad 
      A_0(t') G_{0\,ba}(t',t)
   \>.
\end{eqnarray*}
Here, all functions are to be evaluated using the first order
solutions, $A_0(t)$.

%
%%%%%%%%%%%%%%%%%%%%%%%%%%%%%%%%%%%%%%%%%%%%%%%%%%%%%%%%%%%%%%%%%%%
% 
\subsection{Effective Action approach}

A second method of evaluation of the order $1/N$ corrections is to use
a Legendre transformation to find the effective action
$\Gamma[A,\phi_k]$
\begin{equation}
   \Gamma[A,\phi] = W[J,j]
      - \int_{\C} {\rm d}t
        \left\{ J(t) A(t) + \sum_{a=1}^{N} j_a(t) \phi_a(t) \right\}
   \>,
\label{eq:Legendre}
\end{equation}
Using the fact that $S_{{\rm eff}}[A_0 ; J,j]$ is a stationary
point, we find, to order $1/N$,
\begin{eqnarray}
   \Gamma[A,\phi]
   & = & \int_{\C} {\rm d}t \,
     \biggl \{
       - \frac{1}{2}
         A \frac{ {\rm d}^2 }{ {\rm d} t^2 } A
       \ + \ \frac{i}{2N} \sum_{a=1}^{N}
                \ln \left [ G^{-1}_{aa}[A](t,t) \right ] 
   \nonumber \\  & & \quad \quad \quad 
       \ + \ \frac{i}{2 \, N} \ln \left [ D^{-1}[A,\phi](t,t) \biggr ]
     \right \}
   \nonumber \\  & &
      - \int_{\C} {\rm d}t \, \int_{\C} {\rm d}t' \,
        \frac{1}{2} \sum_{a,b=1}^{N}
           \phi_a(t) G^{-1}_{ab}[A](t,t') \phi_b(t')
   \> .
   \nonumber \\
\end{eqnarray}
which is the classical expression plus the trace-log terms.  Here,
$G^{-1}_{ab}[A](t,t')$ and $D^{-1}[A,\phi](t,t')$ are functionals of
the full $A(t)$ and $\phi_a(t)$.  Again, in this case we set
$\phi_a(t) = 0$.  This effective action agrees up to order $1/N$ with
the true Legendre transform of the generating functional.  One can now
ignore where this action came from and directly write non-perturbative
equations for $A$ directly from this action which will agree to order
$1/N$ with the previous equations. However at $N=1$ where we will be
making a numerical comparison with the exact answer, the results are
expected to be quite different.  We would like to know how the two
results for $A$ differ and which is more accurate at modest $N$.

The equation of motion for $A(t)$ which follows from varying the
effective action $\Gamma$ is:
\begin{eqnarray}
   &&
   \left  \{ \frac{ {\rm d}^2 }{ {\rm d} t^2 } +
      \frac{e^2}{iN} \sum_{a=1}^{N}
      {\cal G}_{aa}(t,t)
   \right \} A(t)
   \nonumber \\ && \quad \quad \quad
   \ + \ 
   \frac{1}{N} \int_{\C} {\rm d}t' \, K(t,t') A(t')
   \ = \ 0  
   \>,
\label{eq:Jdef}
\end{eqnarray}
where
\begin{eqnarray}
   && 
   {\cal G}_{ab}(t,t') \ = \
   G_{ab}(t,t') 
   \nonumber \\ && \quad
   + \frac{i}{N} \sum_{c,d=1}^{N} \int_{\C} {\rm d}t_1 \int_{\C} {\rm d}t_2
         \, G_{ac}(t,t_1) \Sigma_{cd}(t_1,t_2) \, G_{db}(t_2,t') 
   \>,
   \nonumber \\
\label{eq:Gfull}
   \\ &&
   \Sigma_{cd}(t_1,t_2) \ = \
      e^2  \delta_{cd}  \delta_{\C} (t_1,t_2) D(t_1,t_2) 
   \nonumber \\ && \quad \quad
      \ - \ 4 e^4  A(t_1) G_{cd}(t_1,t_2) D(t_1,t_2) A(t_2)
   \>,
\label{eq:Sigma}  
   \\ &&
   K(t,t') \ = \
      2 e^4 D(t,t') \  \frac{1}{N}
      \sum_{a,b=1}^{N}
         \left \{
            G_{ab}(t,t') \, G_{ba}(t',t)
         \right \}
   \>.
   \nonumber \\
\label{eq:Keqn}
\end{eqnarray}
The equation for $A$ has to be solved simultaneously with the
equations for $G$ and $D$ which now depend on the full $A$.
To obtain $G$ we now need to solve for the modes $f$ which satisfy:
\begin{equation}
 \left\{ 
      \frac{ {\rm d}^2 }{ {\rm d} t^2 }  +   m^2 + e^2 A^2(t)  
   \right\}  f_{a}(t)  =  0. 
\label{eq:fnew}
\end{equation}
$G$ is again given by:
\begin{eqnarray}
   &&
   G_{ab}(t,t') \ = \
   \nonumber \\ &&
      i \delta_{ab} 
      \left \{ 
         f_{a}(t) f_{a}^{\ast}(t')  \Theta_{\C}(t,t') + 
         f_{a}(t') f_{a}^{\ast}(t)  \Theta_{\C}(t',t)
      \right \}
   \>.
\end{eqnarray}
In terms of the full $A$ and these new $f$ we determine $D$ using the
integral equation (\ref {eq:Dfcn}) where now $D_0$ is determined using
the new $f$ in the equation for $g$.

These equations of motion agree with the previous ones to order
$1/N^2$.  We had hoped that these equations which partially resum
$1/N$ corrections would be more accurate at late times.  However this
turned out not to be the case, and in fact since $A$ becomes unbounded
sooner in this second approach, the late time behavior of this
approach is worse. At $N=1$ both methods agree with the exact answer
for the same amount of time. When they diverge from each other they
also diverge from the exact answer.

%
%%%%%%%%%%%%%%%%%%%%%%%%%%%%%%%%%%%%%%%%%%%%%%%%%%%%%%%%%%%%%%%%%
%
\section{Energy}
\label{sec:IV}

The expectation value of the energy is given by:
\begin{equation}
   E/N \ = \
   \frac{1}{2} \langle \dot{A}^2 \rangle +
   \frac{1}{2} \sum_{a=1}^{N}
      \left  \{
         \langle \dot{\phi_a}^2 \rangle +
         m^2 \langle \phi_a^2 \rangle +
         e^2 \langle A^2 \phi_a^2
         \rangle
      \right \}
   \>.
\label{eq:energy}
\end{equation}
The expectation values are related to the full connected Green's
Functions ${\bar G }$,${\bar D }$ $K^3_{ab}$, $G^4 _{ab}$ by the
following equations:
\begin{eqnarray}
   &&
   \langle A(t)A(t') \rangle \ = \ 
       A (t) A(t') + \frac{1}{i N}  {\bar D}(t,t')
   \\ &&
   \langle \phi_a(t)\phi_b(t') \rangle \ = \ 
       \frac{1}{i}  {\bar G}_{ab}(t,t')
   \\ &&
   \langle \phi_a(t_1) \phi_b(t_2) A(t_3) A(t_4) \rangle 
   \ = \ - \, \frac{1}{i N^2} G^4 _{ab}(t_3,t_4; t_1,t_2)  
   \nonumber \\ && \quad \quad
         \, + \, \frac{1}{i} A (t_3)A(t_4) {\bar G}_{ab}(t_1,t_2)
         \, - \, \frac{1}{N} {\bar D }(t_3,t_4){\bar G}_{ab}(t_1,t_2)  
   \nonumber \\ && \quad \quad
         \, - \, \frac{1}{N} 
                 \left  \{
                    K^3_{ab}(t_1,t_2; t_3) A(t_4) \, + \, 
                    K^3_{ab}(t_1,t_2; t_4) A(t_3)
                 \right \}
   \nonumber \\
\end{eqnarray}

We notice that to order $1/N$ the connected 3 point function $K^3$
contributes but not the connected 4 point function $G^4$.  If we are
using the generating functional approach we can now directly expand
the energy in a power series in $1/N$ by assuming $G^n = G^n_0 +
\frac{1}{N} G^n_1 + ...$ We then obtain:
\begin{eqnarray*}
   &&
   \langle A^2(t) \rangle
   \ = \
   A_0^2(t) \ + \ \frac{2}{N} \, A_0(t) A_1(t)
   \ + \ \frac{1}{i N} \, D(t,t)
   \\ &&
   \langle \dot{A}^2(t) \rangle
   \ = \
   \dot{A}_0^2(t) + \frac{2}{N} \, \dot{A}_0(t) \dot{A}_1(t)
    + \frac{1}{i N} 
         \left . \frac{\partial^2 D(t,t')}
                      {\partial t \, \partial t'}
         \right |_{t=t'}
   \\ &&
   \langle \phi_a^2(t) \rangle
   \ = \
   \frac{1}{i} 
   \left  \{
      G_{0 \, aa}(t,t) + \frac{1}{N} G_{1 \, aa}(t,t)
   \right \}
   \\ &&
   \langle \dot \phi_a^2(t) \rangle
   \ = \ \frac{1}{i} 
   \left  \{
      \left .
         \frac{\partial^2 \, G_{0 \, aa}(t,t')}
              {\partial t \, \partial t'}
      \right |_{t=t'} \!\!\!
    + \frac{1}{N} 
      \left . \frac{\partial^2 G_1(t,t')}
                   {\partial t \, \partial t'}
      \right |_{t=t'}  
   \right \}
   \\ &&
   \langle A^2(t) \, \phi_a^2(t) \rangle
   \ = \ \frac{1}{i} 
   \biggl \{
      A_0^2(t) \, G_{0 \, aa}(t,t)
   \\ && \quad \quad
    + \frac{1}{N} \, A_0^2(t) \, G_{1 \, aa}(t,t)
    + \frac{2}{N} \, A_0(t) A_1(t) \, G_{0 \, aa}(t,t)
   \biggr \}
   \\ && \quad \quad
    - \frac{1}{N} \, G_{0 \, aa}(t,t) D(t,t) 
    - \frac{2}{N} \, A_0(t)  K^3_0(t,t;t) 
   \>.
\end{eqnarray*}
Where at order $1/N$
\begin{eqnarray}
   &&
   K^3_0 (t_1,t_2;t_3) \ = \ 
      - \, 2 e^2 \,
   \nonumber \\ && \quad \quad
      \int_{\C} {\rm d}t' \, G_0(t_1,t') \, G_0(t_2,t') \, 
         D(t_3 ,t') \, A_0(t') 
   \>,
\end{eqnarray}
and $G_1(t,t')$ is the sum of three terms:
\begin{equation}
   G_{1 \, ab}(t,t')
   \ = \
      G_{1 \, ab}^{(a)}(t,t') +
      G_{1 \, ab}^{(b)}(t,t') +
      G_{1 \, ab}^{(c)}(t,t')
\label{eq:G1}
\end{equation}
with
\begin{eqnarray*}
   &&
   G_{1 \, ab}^{(a)}(t,t')
   \ = \
   i e^2 \int_{\C} {\rm d}t_1 \,
         G_{0 \, ab}(t,t_1) \, D(t_1,t_1) \, G_{0 \, ba}(t_1,t') 
   \\ &&
   G_{1 \, ab}^{(b)}(t,t')
   \ = \
   - 4 i e^4 \sum_{c,d=1}^{N}
      \int_{\C} {\rm d}t_1 \, 
         G_{0 \, ac}(t,t_1) A_0(t_1) \,
   \nonumber \\ && \quad \quad 
      \int_{\C} {\rm d}t_2 \,
         G_{0 \, cd}(t_1,t_2) \, D(t_1,t_2) \,
         G_{0 \, db}(t_2,t') A_0(t_2)
   \\ &&
   G_{1 \, ab}^{(c)}(t,t')
   \ = \
   \nonumber \\ && \quad \quad 
   - 2 e^2 \int_{\C} {\rm d}t_1 \,
           G_{0 \, ab}(t,t_1) A_0(t_1) A_1(t_1)
           G_{0 \, ba}(t_1,t') 
   \>.
\end{eqnarray*}
The above lead to the following expression of the energy:
\begin{equation}
   E_Z(t) \ = \ E_0(t) \ + \ \frac{1}{N} E_1(t)
   \>,
\label{eq:energyZ}
\end{equation}
For $E_0$, we find:
\begin{eqnarray}
   &&
   \frac{E_0(t)}{N}
   \ = \
   \frac{1}{2} \dot A_0^2(t)
   \ + \
   \frac{1}{2iN} \sum_{a=1}^{N}
      \biggl \{
         \left . \frac{\partial^2 G_{0 \, aa}(t,t')}
                      {\partial t \, \partial t'} \right |_{t=t'}
   \nonumber \\ && \quad \quad \quad \quad \quad \quad 
         \ + \
         ( m^2 + e^2 A_0^2) \, G_{0 \, aa}(t,t)
      \biggr \}
   \>.
\label{eq:energy0}
\end{eqnarray}
and where
\begin{eqnarray}
   && 
   \frac{E_1(t)}{N}
   \ = \
   \dot A_0(t) \dot A_1(t)
   \ + \
   \frac{1}{2iN} \sum_{a=1}^{N}
      \biggl\{
         \left . \frac{\partial^2 G_{1 \, aa}(t,t') }
                      {\partial t \, \partial t'} \right |_{t=t'}
   \nonumber \\ &&
          + 
         ( m^2 + e^2 A_0^2) \, G_{1 \, aa}(t,t)
          + 
         2 e^2 \, A_0(t) A_1(t) \, G_{0 \, aa}(t,t)
      \biggr\}
   \nonumber \\ &&
   \ + \ \frac{1}{2} 
   \left  \{
      \frac{1}{i} \left . \frac{\partial^2 D(t,t') }
                               {\partial t \, \partial t'} \right |_{t=t'}
      \ - \ \frac{e^2}{N} \sum_{a=1}^{N} G_{0 \, aa}(t,t) D(t,t)
   \right \}
   \nonumber \\ &&
   \ - \
   e^2 \, A_0(t) \, K_0^3(t,t; t)
   \>.
\label{eq:energy1}
\end{eqnarray}

If instead we are using the effective action $\Gamma$ to determine the
equations of motion, then we find the following expressions for the
expectation values of the operators:
\begin{eqnarray}
   &&
   \langle A^2(t) \rangle
   \ = \
   A^2(t) \ + \ \frac{1}{iN} D(t,t)
   \\ &&
   \langle \dot{A}^2(t) \rangle
   \ = \
   \dot{A}^2(t)
   \ + \ \frac{1}{iN}
         \left . \frac{\partial^2 \, D(t,t')}
                      {\partial t \, \partial t'}
         \right |_{t=t'}
   \\ &&
   \langle \phi_a^2(t) \rangle
   \ = \
   \frac{1}{i} {\cal G}_{aa}(t,t)
   \\ &&
   \langle \dot \phi_a^2(t) \rangle
   \ = \
   \frac{1}{i} 
   \left . \frac{\partial^2 \, {\cal G}_{aa}(t,t')}
                {\partial t \, \partial t'}
   \right |_{t=t'}
   \\ &&
   \langle A^2(t) \, \phi_a^2(t) \rangle
   \ = \
   \frac{1}{i} 
   A^2(t) \, {\cal G}_{aa}(t,t)
   \ - \ \frac{1}{N} \, G_{aa}(t,t) D(t,t)
   \nonumber \\ &&  \quad \quad \quad \quad \quad \quad \quad
   \ - \ \frac{2}{N} \, A(t) \, K^3(t,t; t) \>,
\end{eqnarray}
where now we use the full $A$ and not $A_0$ in determining all the
Green's functions in $K^3$.  The energy for the $\Gamma$-method can
then be calculated as
\begin{eqnarray}
   && 
   \frac{E_\Gamma}{N}
   \ = \
   \frac{1}{2} \dot A^2(t)
   \ + \
   \frac{1}{2iN} \sum_{a=1}^{N}
   \biggl \{
      \left . \frac{\partial^2 {\cal G}_{aa}(t,t') }
                {\partial t \, \partial t'} \right |_{t=t'}
   \nonumber \\ &&
       + \ ( m^2 + e^2 A^2) \, {\cal G}_{aa}(t,t) 
   \biggr \}
   \ + \
   \frac{1}{2N} \,
   \biggl \{
      \frac{1}{i} \left . \frac{\partial^2 D(t,t') }
                               {\partial t \, \partial t'} \right |_{t=t'}
   \nonumber \\ &&
      \ - \ \frac{e^2}{N} \sum_{a=1}^{N} G_{aa}(t,t) D(t,t) 
   \biggr \}
   \ - \
   \frac{e^2}{N} \, A(t) \, K^3(t,t; t) \>.
\label{eq:energyG}
\end{eqnarray}
Using the equations of motion, we can show that all expressions for
the energy, Eqs.~(\ref{eq:energy0}, \ref{eq:energyZ},
\ref{eq:energyG}), are time independent, so that the energy is
conserved for both the perturbative calculation as well as the
effective action method.

%
%%%%%%%%%%%%%%%%%%%%%%%%%%%%%%%%%%%%%%%%%%%%%%%%%%%%%%%%%%%%%%%%%%%
%
\section{Numerical results}
\label{sec:V}

In order to find numerical solutions to these equations, we expand all
functions in Chebyshev polynomials, and solve the resulting finite set
of equations.  In most of our calculations, we used a set of 32
polynomials.  Details of this calculation are given in Appendix
\ref{sec:B}.

In order to make contact with our previous lowest order $N= \infty$
results we will use the set of initial conditions ($e=0.3$, $E=1$,
$m=1$) given in reference~\cite{ref:CDHR}.  Time is measured in units
of $1/m$.  We would first like to compare the two methods at large but
finite $N$ starting off with Gaussian initial data for the wave
function.  In a mean field approximation, such as lowest order large
$N$, a wave function which starts out as a Gaussian remains Gaussian.
Thus in mean field theory the only physical measurable quantities are
$ \langle A(t) \rangle$, $\langle (A(t) - \langle A(t) \rangle)^2
\rangle = D(t,t)/i$, and $\langle \phi^2(t) \rangle = {\cal
G}(t,t)/i$.  It is just these three moments that we will plot here to
make contact with mean field results. (At infinite $N$, $D$ is also
zero, but it is finite in a Hartree approximation, as discussed
in~\cite{ref:CDHR}).  We will denote the straightforward use of the
$1/N$ expansion from the path integral as the $Z$-method and the
results coming from the effective action approach, the
$\Gamma$-method.  

\begin{figure}
   \centering
   \epsfxsize = 3.0in
   \subfigure[$N=100$.]
             {\epsfbox{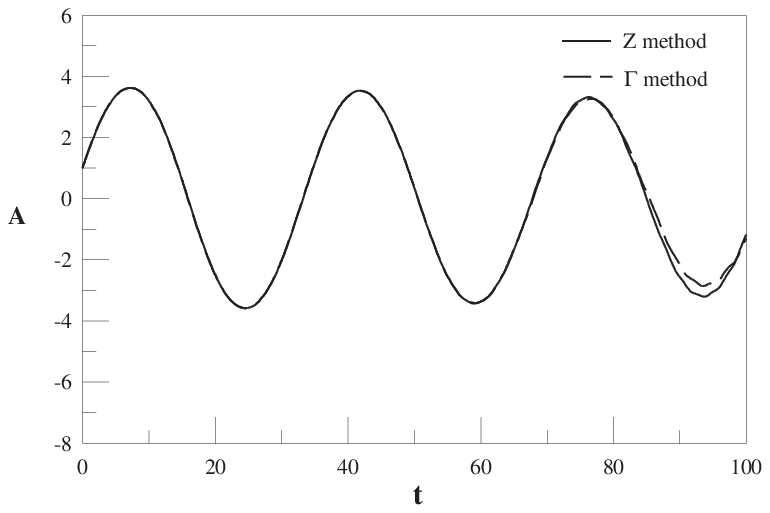}}
   \subfigure[$N=16$.]
             {\epsfbox{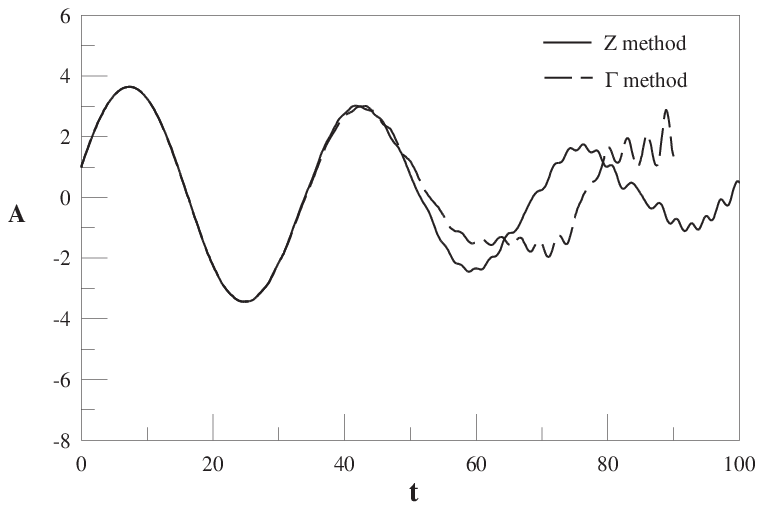}}
   \subfigure[$N=8$.]
             {\epsfbox{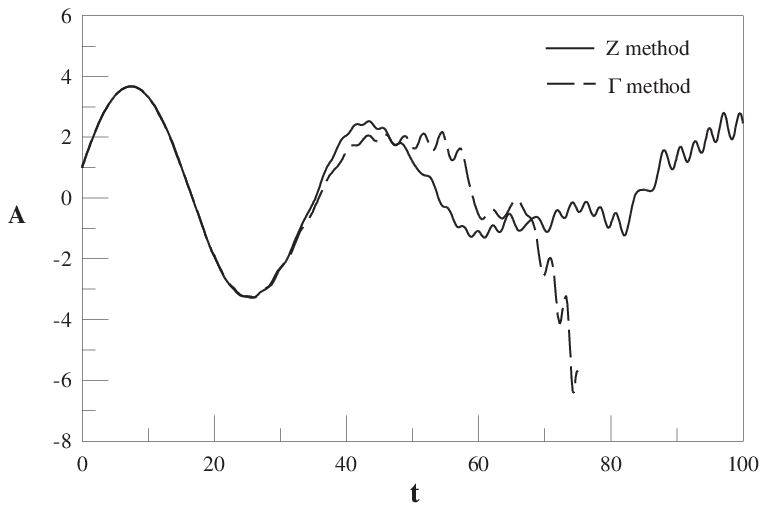}}
   \caption{Time evolution of $A(t)$ for $N=8$, 16, and 100,
            using the $Z$-method and the $\Gamma$-method.}
   \label{fig:Aabc}    
\end{figure}

In Fig.~\ref{fig:Aabc}, we plot $\langle A(t) \rangle$ for $N= 100,
16$, and 8.  We notice that as we reduce $N$ the two different methods
start diverging after one period.  We also find that the second method
which is based on the effective action has corrections which become
unbounded at late times which leads us to believe that the second
method is less stable.  Both methods seem to have secular behavior
({\em i.e.} corrections which grow as a power in time).

In Fig.~\ref{fig:Gabc}, we show results for $\langle \phi^2(t)
\rangle = {\cal G}(t,t)/i$ for $N= 100, 16$, and 8.  We notice that at
late times, this can go negative.  This is a result of the fact that
the operator $\phi$ has a $1/N$ expansion $\phi = \phi_0 +
\frac{1}{N}\phi_1 + \cdots$. So that the positive definite width of
the wave packet is $\langle (\phi_0 + \frac{1}{N}\phi_1)^2 \rangle$
which also includes $1/N^2$ corrections.  So once the $1/N$ correction
becomes as large as the lowest order term, the expansion breaks down
and keeping terms only to order $1/N$ in the expansion of the
expectation value can lead to negative results.  This negative result
occurs after the two methods diverge which is an indication of when
the expansion is breaking down.

\begin{figure}
   \epsfxsize = 3.0in
   \centerline{\epsfbox{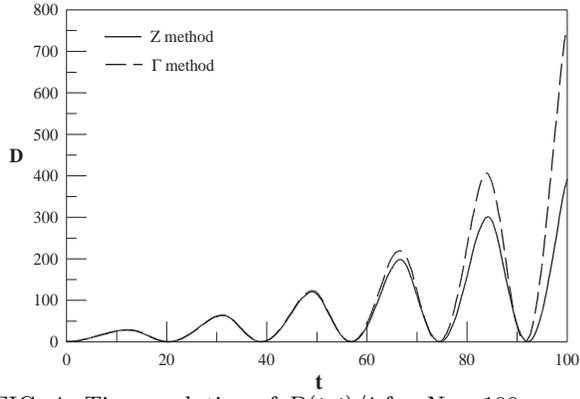}}
   \caption{Time evolution of $D(t,t)/i$ for $N=100$, as computed using the
            $Z$-method and the $\Gamma$-method.}
   \label{fig:N100_Dtt}
\end{figure}

\begin{figure}
   \epsfxsize = 3.0in
   \centerline{\epsfbox{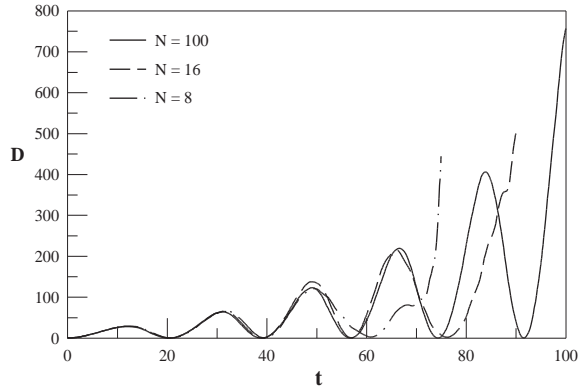}}
   \caption{Time evolution of $D(t,t)/i$ for $N=8,16,100$, 
            as computed using the $\Gamma$-method.}
\label{fig:Nvar_Dtt}
\end{figure}

Now let us look at the effective width of the wave function of the $A$
oscillator. For the $Z$-method, we have that the correlation function
$D(t,t')$, defined by Eq.~(\ref{eq:Dfcn}), is independent of $N$.
However, the width determined in the effective action approach has
$1/N$ corrections due to the implicit dependence of $A(t)$ on $N$.
First, in Fig.~\ref{fig:N100_Dtt}, we compare the two approaches for
calculating $D(t,t)/i$ at $N=100$.  In Fig.~\ref{fig:Nvar_Dtt}, we
show the $N$ dependence of $D(t,t)/i$ using the $\Gamma$-method at $N=
100, 16$, and 8.  From Fig.~\ref{fig:Nvar_Dtt} we see that when $t
\geq 5$ the $1/N^2$ effects are starting to appear.
 
\begin{figure}
   \epsfxsize = 3.0in
   \centerline{\epsfbox{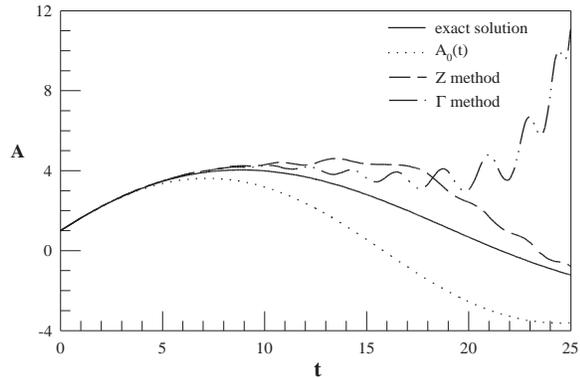}}
   \caption{Time evolution of $A(t)$ for $N=1$, as computed using the
            $Z$-method and the $\Gamma$-method, compared with
            the exact solution.}
   \label{fig:A_ZGprn}
\end{figure}

For the case $N=1$, it is possible to compare the results of both
expansions with an exact numerical simulation of the Schr\"odinger
equation which we obtained in reference~\cite{ref:CDHR}.
Fig.~\ref{fig:A_ZGprn} shows the time evolution of $A(t)$ as computed
using both the $\Gamma$- and the $Z$-method, compared with the exact
solution for $0 < t < 25$. For comparison we also include the lowest
order in large $N$ result.  Both methods agree with each other and
with the exact result for $t < 8$.  They are accurate for at least
twice the time scale as $A_0(t)$, the first order large $N$ result. It
is gratifying to see that the two solutions diverge from each other
approximately the same time as when they diverge from the exact
solution so that the divergence of the two approaches which signals
the onset of $1/N^2$ corrections sets the time scale for the accuracy
of the result even for~$N=1$.

\begin{figure}
   \epsfxsize = 3.0in
   \centerline{\epsfbox{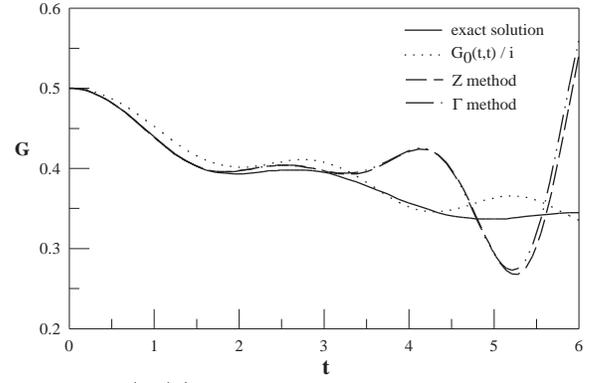}}
   \caption{${\cal G}_{aa}(t,t)/i$ for the first and
            second order large $N$ approximations as a function of
            time.}
   \label{fig:G_Gprn}
\end{figure}

In Fig.~\ref{fig:G_Gprn}, we plot the effective width of the $\phi$
oscillators $\langle (\phi(t) - \langle \phi(t) \rangle)^2 \rangle =
{\cal G}_{aa}(t,t)/i $ using both methods and compare these results to
the exact solution for $N=1$ as well as the lowest order result.  Here
we notice a marked improvement of the $1/N$ corrected results from the
lowest order ones.  We find again that the time at which the two
methods start diverging from one another they also diverge from the
exact answer.

\begin{figure}
   \epsfxsize = 3.0in
   \centerline{\epsfbox{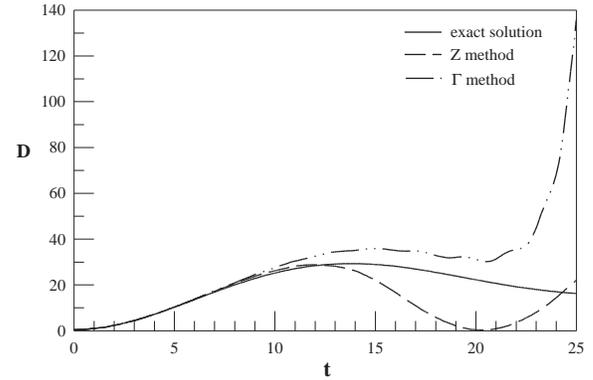}}
   \caption{Comparison of exact numerical simulation of $D(t,t)/i$ with
            the value computed using the $Z$ and $\Gamma$ methods.}
\label{fig:D_ZGprn}
\end{figure}

In Fig.~\ref{fig:D_ZGprn}, we plot the effective width of the $A$
oscillator $D(t,t)/i$ for the $Z$- and $\Gamma$-methods, as a function
of time.  In this case the lowest order in $1/N$ is equivalent to a
delta function width and so did not give a prediction.  Here again we
find the two solutions diverge from one another just when the
approximation breaks down.

What we can conclude from the above results are that $1/N^2$
corrections become important rather early in this particular time
evolution problem. The period of agreement of our two methods is
approximately the period when they give accurate results for the time
evolution.  We find at $N=1$ that the width of the $\phi$ oscillator
is much better described when we add the $1/N$ corrections. Also the
time evolution of the $A$ oscillator is now accurate for about twice
the time period found previously at lowest order.

In our calculations, we were able to verify energy conservation to one
part in $10^4$.  In spite of this, when the $1/N$ expansion breaks
down it fails to preserve the positivity of certain expectation
values.  If we look at Eq.~(\ref{eq:Dfcn}) for $D(t,t)/i$, which is
the positive definite expectation value $ \langle ( A(t) - \langle
A(t) \rangle )^2 \rangle$, we notice that this is an integral equation
which sums all the bubbles and thus correctly takes into account the
shift in the frequency of the quantum fluctuations of the $A$
oscillator.  However in the equation for $\langle A(t) \rangle$
itself, Eq.~(\ref{eq:Jdef}), the term which is the time dependent mass
of the oscillator, namely $\langle \phi^2(t) \rangle$, which should
also be positive definite, is given by Eq.~(\ref{eq:Gfull}).  We
notice that Eq.~(\ref{eq:Gfull}) is not an integral equation but only
the leading term in a $1/N$ expanded Green's function. This quantity
need not be (and is not) positive definite.  We can understand how
this can happen in an example.  In the vacuum state $\langle \phi^2(t)
\rangle = G(t,t)/i$ is time independent and is positive definite.  One
of the effects of the interactions is to renormalize the mass (change
the frequency) of the oscillator.  So consider the toy problem:
\begin{eqnarray*} 
   G(t,t)
   & = &
   \frac{1}{2\pi} \int_{-\infty}^{\infty} \, 
     \frac{ d\omega }
           { \left( \omega^2 + m_0^2 + \frac{e^2}{N} m_1^2 \right) }
   \nonumber \\ & = &
   \frac{1}{2} \left(  m_0^2 + \frac{e^2}{N} m_1^2 \right)^{-1/2}  
\end{eqnarray*} 
where we have kept only the $e^2 /N$ correction to the mass shift of
the $\phi$ oscillator.  Now $m_1^2$ is positive since we have assumed
the interaction is repulsive.  Again reexpanding $G$ we get
\begin{displaymath}
   G(t,t) = \frac{1}{2 m_0} 
      \left  ( 
         1 - \frac{e^2 m_1^2}{2 N m_0^2} 
      \right ) 
\end{displaymath}
So we see that if $m_1^2$ is large enough, then this quantity which is
positive definite can appear negative when reexpanded in $1/N$.

This problem can also be a source of secular terms.  For example
consider the fact that the frequency of the classical $A$ oscillator
is shifted by $1/N$ corrections so that one has
\begin{equation}
   A(t) = A(0) \cos 
      \left  ( 
         \omega_0 t + \frac{\omega_1}{N} t 
      \right )
\end{equation}
If however the actual $1/N$ expansion for the oscillator instead gave the
first two terms in the Taylor series we would get:
\begin{equation}
   A(t) = A(0) [ \cos (\omega_0 t) - 
      \frac{\omega_1 t}{N} \sin (\omega_0 t) ]
\end{equation}
Using this expression would lead to blow up of $A(t)$ in a time scale
$N/\omega_1$.  This type of behavior is found in our numerical
simulations.  One way of solving the secular problem as well as
guaranteeing the positivity of expectation values is by appropriately
summing the series.  That is we replace (\ref{eq:Gfull}) by:
\begin{eqnarray}
   &&
   {\cal G}_{ab}(t,t') \ = \ G_{ab}(t,t') 
   \nonumber \\ && 
   \ + \ \frac{i}{N} \sum_{c,d=1}^{N} 
         \int_{\C} {\rm d}t_1 \int_{\C} {\rm d}t_2 \, 
            G_{ac}(t,t_1) \, \Sigma_{cd}(t_1,t_2) \, 
            {\cal G}_{db}(t_2,t')
   \>.
   \nonumber \\
   \label{eq:Gfull2} 
\end{eqnarray}
We then have to modify appropriately Eqs.~(\ref{eq:D0}), (\ref{eq:Pi0}), 
(\ref{eq:Sigma}) and (\ref{eq:Keqn}) in order to guarantee 
energy conservation again.

These modifications only change the results at order $1/N^2$ but are
obviously very important because of the early breakdown of the naive
methods. However in this paper we did not want to go beyond what one
obtains directly using the two direct approaches to the $1/N$
expansion and will discuss the resummed theory in a separate paper.

\begin{figure}
   \epsfxsize = 3.0in
   \centerline{\epsfbox{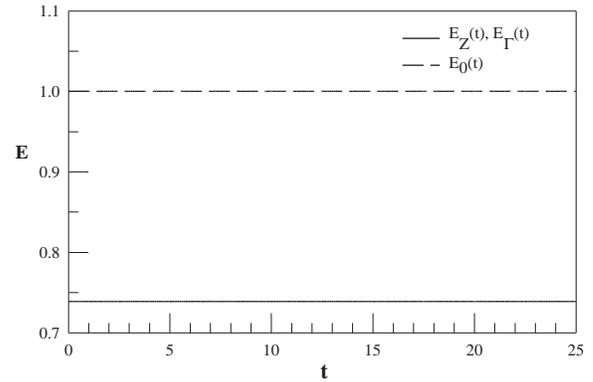}}
   \caption{The energy $E_Z = E_0 + E_1 / N$, computed using the $Z$-method,  
            and $E_{\Gamma}$, using the $\Gamma$-method. }
\label{fig:Energy}
\end{figure}

This problem with expectation values becoming unbounded is more acute
in the $\Gamma$ method.  In Fig.~\ref{fig:Energy} we show that energy
is conserved.  However in Figs.~\ref{fig:E_Gcomp} and
\ref{fig:E_Zcomp} we show the above phenomena that several of the
individual contributions to the energy become negative if we keep
terms only up to $1/N$.

The appearance of secular terms in a perturbation expansion is quite
well known~\cite{ref:Bender}. As an example for the classical
anharmonic oscillator
\begin{displaymath}
   \frac{d^2 y}{d t^2} + y + g y^3 = 0 \>,
\end{displaymath}
if we assume a solution of the form
\begin{displaymath}
   y = y_0(t) + g y_1(t) \>,
\end{displaymath}
and perform a perturbation series in the coupling constant $g$,
secular terms linear in $t$ appear in $y_1(t)$. One can see by doing a
large-$N$ expansion in the Schr\"odinger picture, that one can avoid the
secular problem in a simple way.  If we solve the Schr\"odinger
equation at large $N$ in terms of the eigenfunctions and eigenvalues we
find:
\begin{equation}
   \Psi(\phi,A, t) = \sum_m c_m \Psi_m(\phi,A) e^{-i E_m t }
\end{equation}
where the $\Psi_m(\phi,A)$ are the solutions of the time independent
Schr\"odinger equation. We can do a large $N$ expansion for the
eigenfunctions and eigenvalues separately.  We then find that
$\Psi_m(\phi,A)$ has a Taylor series in $1/\sqrt{N}$ and $E_m$ has a
Taylor series in $1/N$.  As long as we do not reexpand the
exponentials having the time dependence ({\em i.e.} keep $ \exp \{ - i
[ E_m^0 t + \frac{1}{N} E_m^1 t \cdots] \}$ ) then there are no
secular terms.  However if we reexpand the exponentials and keep only
terms of order $1/N$, secular terms proportional to powers of $t$ will
appear.  We believe that this might very well be occuring in the large
$N$ expansion based on the CTP formalism.

%
%%%%%%%%%%%%%%%%%%%%%%%%%%%%%%%%%%%%%%%%%%%%%%%%%%%%%%%%%%%%%%%%%%%%
%
\section{Conclusions}
\label{sec:VII}

We have presented two methods for calculating the $1/N$ corrections to
the time evolution of a system of $N+1$ oscillators using Schwinger's
CTP formalism. These two methods differ by terms of order $1/N^2$ so
that the divergence of results is an indication of the time during
which keeping terms up to order $1/N$ is accurate. This was verified
by comparing with direct numerical simulation of the Schr\"odinger
equation for the case $N=1$. We found that for $N=1$, keeping the
$1/N$ corrections allows us to extend the range of time for which the
expectations values track the exact answer significantly. This was
most noticable for the the quantity which describes the width of the
$\phi$ wave function.

In performing these numerical simulations, we found some shortcomings
in the $1/N$ expansion.  It does not guarantee the boundedness of
various expectation values, even though energy is exactly
conserved. As a result of this, we expect that we are seeing secular
behavior in our expansion which needs to be cured.  One way of curing
this problem is to instead look at a $1/N$ expansion for the
eigenfunctions and eigenvalues of the Schr\"odinger equation instead
of working with the CTP formalism.  Another is to resum the Green's
function for the $\phi$ oscillator and modify the equations for the
other variables appropriately so that energy is conserved.  These two
ideas will be discussed in a future work.  

In conclusion, by comparing our two direct approaches to performing a
$1/N$ expansion, one using the generating functional $Z$ and one the
effective action $\Gamma$, we found that the direct perturbation
theory in $1/N$ from $Z$ gave results that were bounded for longer
times.  However having both methods, allowed us to determine when
$1/N^2$ corrections became important.  Thus having both methods will
be quite useful in simulating field theories, when there will be no
exact solution to compare with.  Solving the secular problem we found
here will be discussed elsewhere.

%
%%%%%%%%%%%%%%%%%%%%%%%%%%%%%%%%%%%%%%%%%%%%%%%%%%%%%%%%%%%%%%%%%%%
%
\acknowledgments

The large scale numerical calculation for the exact solutions was
performed by Salman Habib on the CM5 at the Advanced Computing
Laboratory, Los Alamos National Laboratory and we would like to thank
him for his continuous help.  The authors also acknowledge helpful
conversations with Emil Mottola, Yuval Kluger, and Melissa Lampert.
UNH gratefully acknowledges support by the U.S. Department of Energy
(DE-FG02-88ER40410).

%
%%%%%%%%%%%%%%%%%%%%%%%%%%%%%%%%%%%%%%%%%%%%%%%%%%%%%%%%%%%%%%%%%
%
\appendix

%
%%%%%%%%%%%%%%%%%%%%%%%%%%%%%%%%%%%%%%%%%%%%%%%%%%%%%%%%%%%%%%%%%
%
\section{Equation for $D(t,t')$}
\label{sec:A}

In this appendix we discuss how we approach the problem of finding the
solution of eq. (\ref{eq:Dfcn})
\begin{eqnarray}
   &&
   D(t,t') \ = \ D_0(t,t') 
   \nonumber \\ && \quad
   \ -  \
     \int_{\C} {\rm d}t_1 \int_{\C} {\rm d}t_2 \, D_0(t,t_1)
                                               \, \Pi(t_1,t_2) D(t_2,t')
   \>.
\label{eq:Dfulleqn}
\end{eqnarray}
We first note that the causal Green's functions are symmetric in the
sense that $ {\cal A}_{>}(t,t') = {\cal A}_{<}(t',t) $, and obey the
additional condition
\begin{equation}
   {\cal A}_{>,<}(t,t') \ = \
       - \ {\cal A}^{\ast}_{<,>}(t,t') \ = \ {\cal A}_{<,>}(t',t)
   \>.
\label{eq:Asym}
\end{equation}
The last equation gives
\begin{eqnarray}
   \real \{ {\cal A}_{>}(t,t') \} & = &
          - \ \real \{ {\cal A}_{<}(t,t') \}
\label{eq:Are}
   \\
   \imag \{ {\cal A}_{>}(t,t') \} & = &
          - \ \imag \{ {\cal A}_{<}(t,t') \}
   \>,
\label{eq:Aim}
\end{eqnarray}
or
\begin{eqnarray}
   {\cal A}_{>}(t,t') - {\cal A}_{<}^{\ast}(t,t')
   & = & 2 \, \real \{ {\cal A}_{>}(t,t') \}
\label{eq:Adif}
   \\
   {\cal A}_{>}(t,t') + {\cal A}_{<}^{\ast}(t,t')
   & = & 2 \, \imag \{ {\cal A}_{>}(t,t') \}
   \>.
\label{eq:Asum}
\end{eqnarray}
With these equations in mind, the causal Green's fumction ${\cal
A}(t,t')$ is fully determined by the component ${\cal A}_{>}(t,t') =
\real \{ {\cal A}_{>}(t,t') \} + i \ \imag \{ {\cal A}_{>}(t,t') \}$,
and we need to evaluate the function only for $t' \leq t$.

Directing our attention now to the calculation of $D(t,t')$, it is
convenient to introduce
\begin{equation}
   Q(t,t') \ = \
     \int_{\C} {\rm d}t'' \, D_0(t,t'') \, \Pi(t'',t')
   \>.
\label{eq:dQ}
\end{equation}
Note that even though the functions $D_0(t,t')$ and $\Pi(t,t')$
satisfy the properties of the Green's functions listed above, the new
function $Q(t,t')$ satisfies only some, i.e.
\begin{equation}
   Q_{>,<}(t,t') \ = \ - \ Q^{\ast}_{<,>}(t,t') \ \neq \ Q_{<,>}(t',t)
   \>.
\label{eq:Qsym}
\end{equation}
This does not prevent the Green's function $D(t,t')$ to recover all
the desired properties after one more CTP integration.  In practice,
this translates into the fact that we have to actually calculate the
function $Q_>(t,t')$ for all moments $t$ and $t'$, whereas the other
causal functions need only some half of the data.

Then, eq. (\ref{eq:Dfulleqn}) becomes
\begin{equation}
   D(t,t') \, = \, D_0(t,t') \ - \
     \int_{\C} {\rm d}t'' \, Q(t,t'') \, D(t'',t')
   \>,
\label{eq:DQeqn}
\end{equation}
or
\begin{eqnarray}
   &&
   D_{>}(t,t') \ = \ D_{0 \, >}(t,t')
   \nonumber \\ && \quad
     \ - \ \int_{0}^{t} {\rm d}t'' \,
     \left[ Q_{>}(t,t'') - Q_{<}(t,t'') \right ] \, D_{>}(t'',t')
   \nonumber \\ && \quad
     \ + \ \int_{0}^{t'} {\rm d}t'' \,
     Q_{>}(t,t'')\, \left[ D_{>}(t'',t') - D_{<}(t'',t') \right ]
\label{eq:DQbig}
   \\ &&
   D_{<}(t,t') \ = \ D_{0 \, <}(t,t')
   \nonumber \\ && \quad
     \ - \ \int_{0}^{t} {\rm d}t'' \,
     \left[ Q_{>}(t,t'') - Q_{<}(t,t'') \right ] \, D_{<}(t'',t')
   \nonumber \\ && \quad
     \ + \ \int_{0}^{t'} {\rm d}t'' \,
     Q_{<}(t,t'')\, \left[ D_{>}(t'',t') - D_{<}(t'',t') \right ]
   \>.
\label{eq:DQles}
\end{eqnarray}
Using the properties (\ref{eq:Asym}) - (\ref{eq:Asum}), we see that
eq. (\ref{eq:DQles}) is redundant, and we can write
eq. (\ref{eq:DQbig}) as
\begin{eqnarray}
   &&
   D_{>}(t,t') \ = \ D_{0 \, >}(t,t')
   \nonumber \\ && \quad
     \ - \ 2 \ \int_{0}^{t} {\rm d}t'' \,
     \real \{ Q_{>}(t,t'') \} \ D_{>}(t'',t')
   \nonumber \\ && \quad
     \ + \ 2 \ \int_{0}^{t'} {\rm d}t'' \,
     Q_{>}(t,t'') \ \real \{ D_{>}(t'',t') \}
   \>.
\label{eq:DQbig0}
\end{eqnarray}
We separate now the {\em real} and the {\em imaginary} part of
(\ref{eq:DQbig0}) and obtain the system of integral equations
\begin{eqnarray}
   &&
   \real \{ D_{>}(t,t') \}
   \ = \
   \real \{ D_{0 \, >}(t,t') \}
   \nonumber \\ && \quad 
   \ - \ 2 \ \int_{0}^{t} {\rm d}t'' \,
                           \real \{ Q_{>}(t,t'') \} \
                           \real \{ D_{>}(t'',t') \}
   \nonumber \\ && \quad 
   \ + \ 2 \ \int_{0}^{t'} {\rm d}t'' \,
                            \real \{ Q_{>}(t,t'') \} \
                            \real \{ D_{>}(t'',t') \}
\label{eq:ReDQbig0}
   \\ &&
   \imag \{ D_{>}(t,t') \}
   \ = \
   \imag \{ D_{0 \, >}(t,t') \}
   \nonumber \\ && \quad 
   \ - \ 2 \ \int_{0}^{t} {\rm d}t'' \,
                           \real \{ Q_{>}(t,t'') \} \
                           \imag \{ D_{>}(t'',t') \}
   \nonumber \\ && \quad 
   \ + \ 2 \ \int_{0}^{t'} {\rm d}t'' \,
                            \imag \{ Q_{>}(t,t'') \} \
                            \real \{ D_{>}(t'',t') \}
   \>.
\label{eq:ImDQbig0}
\end{eqnarray}
The above system of equations has to be solved for $t' \leq t$.
Notice that the two equations are independent, which allows us to
solve first eq. (\ref{eq:ReDQbig0}) for the {\em real} part of
$D_{>}(t,t')$, and then use this result to derive the {\em imaginary}
part of $D_{>}(t,t')$ from equation (\ref{eq:ImDQbig0}).

%
%%%%%%%%%%%%%%%%%%%%%%%%%%%%%%%%%%%%%%%%%%%%%%%%%%%%%%%%%%%%%%%%%
%
\section{Numerical methods}
\label{sec:B}

Our numerical technique involves the expansion of all the unknown
functions, $f(t)$, $g(t)$, $A(t)$, $D(t,t')$, in a Chebyshev
polynomial basis.  We follow a method developed by
El-gendy~\cite{ref:Elgendy} and have applied it to combined
differential and integrals equations of the type we have here.  We use
the same Chebyshev expansion methods for solving the Green's function
equation for $D(t,t')$, in the CTP formalism, as explained in
appendix~\ref{sec:A}.  In addition, we divide the time up into small
blocks, moving along block by block.

Chebyshev polynomials of the first kind of degree {\em n} are defined
by,
\begin{equation}
   T_n(x) \ = \ \cos(n \, \arccos \, x)
   \> .
   \label{eq:cheby_def}
\end{equation}
We define the grid in the interval $[-1,1]$ by choosing the positions:
\begin{equation}
   \tilde{x}_j \ = \ \cos{\frac{j \, \pi}{N}}
      \qquad j = 0,1,\ldots,N
   \>.
\label{eq:grid}
\end{equation}
Then, the Chebyshev polynomials satisfy a discrete orthogonality
relation of the form
\begin{equation}
   \sum_{k=0}^N \ T_i(\tilde{x}_k) T_j(\tilde{x}_k)
   \ = \
   \beta_i \ \delta_{i \, j}
   \>,
\end{equation}
where the constants $\beta_i$ are
\begin{equation}
   \beta_i
   \ = \
   \left \lbrace
         \begin{array}{lll}
            \displaystyle{\frac{N}{2}} \, ,  & {} & i \neq 0,N \\ \\
            N \, ,                           & {} & i = 0,N
         \end{array}
   \right .
   \>.
\end{equation}
An arbitrary function $f(x)$ can be approximated in the interval
[-1,1] by the formula
\begin{equation}
   f(x)
   \ = \
   \sum_{j=0}^{N} {\rm {}''} \
       a_j \ T_j(x)
   \>,
\label{eq:f_approx}
\end{equation}
We denote $f_k = f(\tilde{x}_k)$.  The coefficients $a_j$ are defined by
\begin{equation}
   a_j \ = \
   \frac{2}{N} \,
   \sum_{k=0}^N {\rm {}''} \
       f(\tilde{x}_k) T_j(\tilde{x}_k) \ , \ \ \ \ j = 0,\ldots,N
\label{eq:coeff}
\end{equation}
and the summation symbol with double primes denotes a sum with first
and last terms halved.  As shown by El-gendy, we never have to
actually compute these expansion coefficients $a_j$.  Instead, we find
$f_j$ directly.  The advantage of the Chebyshev expansion method is
that (\ref{eq:f_approx}), which is an expansion in a {\em finite} set,
is exact for $x = \tilde{x}_j, j = 0,\ldots,N$.

We can approximate the calculation of the indefinite integral
$\int_{-1}^x \ f(t) \, dt$ by
\begin{equation}
   I(x)
   \ = \
   \int_{-1}^x \ f(t) \, dt
   \ = \
   \sum_{j=0}^{N} {\rm {}''} \
       a_j \ \int_{-1}^x \ T_j(t) \, dt
   \>.
\label{eq:f_indefinite}
\end{equation}
where the integral $\int_{-1}^x \ T_j(t) \, dt$ is given by
\begin{equation}
   \left \lbrace
      \begin{array}{ll}
         \displaystyle{
         \frac{T_{j+1}(x)}{2(j+1)} - \frac{T_{j-1}(x)}{2(j-1)} 
         + \frac{(-)^{j+1}}{j^2-1}         
         } 
       & {\rm if} \ j \geq 1 \\ \\
         \frac{1}{4} \left [ T_2(x) - 1 \right ] 
       & {\rm if} \ j = 1 \\ \\
         T_1(x) + 1 
       & {\rm if} \ j = 0
      \end{array}
   \right .
   \>.
\end{equation}
When the variable $x$ takes the values of the grid points
(\ref{eq:grid}), we can write eq.~(\ref{eq:f_indefinite}) in a matrix
format, like
\begin{equation}
   I_i
   \ = \ I(\tilde{x}_i)
   \ = \ \sum_{j=0}^{N} {\rm {}''} \ B^{[-1,1]}_{ij} \ f_j
   \>,
\label{eq:Beqn}
\end{equation}
where $B^{[-1,1]}$ is a square matrix of order (N+1), whose elements
$B^{[-1,1]}_{ij}$ are equal to
\begin{eqnarray}
   &&
   \frac{1}{N^2} T_N(\tilde{x}_i) T_{N-1}(\tilde{x}_j)
    + 
   \frac{1}{2N (N+1)} T_{N+1}(\tilde{x}_i) T_N(\tilde{x}_j)
   \nonumber \\ &&
   \sum^{N}_{\scriptstyle{k=0 \, (k\neq1)}} \! \! \! \! \! {\rm {}''} \ 
       \frac{2}{N} \frac{(-)^{k+1}}{k^2-1} 
       T_0(\tilde{x}_i) T_k(\tilde{x}_j)
    - \frac{1}{2 N} T_0(\tilde{x}_i) T_1(\tilde{x}_j)
   \nonumber \\ &&
    + \frac{1}{N (N-1)} T_{N-1}(\tilde{x}_i) 
               [ T_{N-2}(\tilde{x}_j) - \frac{1}{2} T_N(\tilde{x}_j) ]
   \nonumber \\ &&
    + \sum^{N-2}_{k=1} \
       \frac{1}{k N} T_k(\tilde{x}_i) 
                        [ T_{k-1}(\tilde{x}_j) - T_{k+1}(\tilde{x}_j) ]
   \>.
\label{eq:B_mat}
\end{eqnarray}

In a similar manner, the derivative $df(x)/dx$ is approximated by
\begin{equation}
   D(x)
   \ = \ \frac{d}{dx} \ f(x)
   \ = \ \sum_{j=0}^{N} {\rm {}''} \
         a_j \ {T'}_j(x)
   \>.
\label{eq:f_derivative}
\end{equation}
where ${T'}_j(x)$ is the derivative of the Chebyshev polynomial of the
first kind of degree $j$.  By inserting the expression of the $a_j$
coefficients, and evaluating the derivative at $x = \tilde{x}_k$ we
obtain a matrix equation,
\begin{equation}
   D_i
   \ = \ D(\tilde{x}_i)
   \ = \ \sum_{j=0}^{N} {\rm {}''} \ \tilde{B}^{[-1,1]}_{ij} \ f_j
   \>.
\label{eq:BTeqn}
\end{equation}
Here $\tilde{B}^{[-1,1]}$ is given by:
\begin{equation}
   \tilde{B}^{[-1,1]}_{ij}
   \ = \
   \frac{2}{N}
   \sum_{k=0}^N {\rm {}''} \
       {T'}_k(\tilde{x}_i) \ T_k(\tilde{x}_j)
   \>.
\label{eq:Btil_mat}
\end{equation}
Note that the approximations (\ref{eq:Beqn}) and (\ref{eq:BTeqn})
are {\em exact} for $x = \tilde{x}_j, j = 0,\ldots,N$.

Finally, the approximations (\ref{eq:f_approx}, \ref{eq:Beqn},
\ref{eq:BTeqn}) can be generalized by allowing the range of the
approximation to be between two arbitrary limits {\em a} and {\em b},
instead of just -1 to 1.  This is done performing the change of
variable
\begin{equation}
   x \ \rightarrow \
   y \ \equiv \
   \frac{x - \frac{1}{2}(b+a)}{\frac{1}{2}(b-a)}
   \>.
\label{eq:x_to_y}
\end{equation}
As a consequence, the matrices $B$ and $\tilde{B}$ become
\begin{eqnarray}
   B^{[a,b]}
      & = & \frac{b-a}{2} \ B^{[-1,1]}
   \\
   \tilde{B}^{[a,b]}
      & = & \frac{2}{b-a} \ \tilde{B}^{[-1,1]}
\end{eqnarray}
and the Chebyshev polynomials in eq. (\ref{eq:f_approx}) become now
functions of the variable {\em y}.  The matrices $\left [ f \right ]$,
$\left [ \int_{-1}^x \ f(t) \, dt \right ]$ and $\left [ df(x)/dx
\right ]$ will give then the value of the function $f(x)$, its
integral and derivative, at the coordinates $x_k = y_k (b-a) / 2 +
(b+a) / 2$, where $y_k$ are the (N+1) coordinates given by
eq. (\ref{eq:grid}).

%
%%%%%%%%%%%%%%%%%%%%%%%%%%%%%%%%%%%%%%%%%%%%%%%%%%%%%%%%%%%%%%%%%
%

\newpage

%%%%%%%%%%%%%%%%%%%%%%%%%%%%%%%%%%%%%%%%%%%%%%%%%%%%%%%%%%%%%%
%
% Figures:
%

% Fig. 1: CTP path

% Fig. 2a: N100_GZ.eps
% Fig. 2b: N16_GZ.eps
% Fig. 2c: N8_GZ.eps

% Fig. 3a: N100_Gtt.eps
% Fig. 3b: N16_Gtt.eps
% Fig. 3c: N8_Gtt.eps

\begin{figure}
   \centering
   \epsfxsize = 3.0in
   \subfigure[$N=100$.]
             {\epsfbox{fig3a_bm.eps}}
   \subfigure[$N=16$.]
             {\epsfbox{fig3b_bm.eps}}
   \subfigure[$N=8$.]
             {\epsfbox{fig3c_bm.eps}}
   \caption{Time evolution of ${\cal G}(t,t)/i$ for $N= 8$, 16, and 
      100, using the $Z$-method and the $\Gamma$-method.}
   \label{fig:Gabc}
\end{figure}

% Fig. 4: N100_Dtt.eps

% Fig. 5: Nvar_Dtt.eps

% Fig. 6: A_ZGprn.eps

% Fig. 7: G_Gprn.eps

% Fig. 8: D_ZGprn.eps

% Fig. 9: Energy.eps

% Fig. 10: E_Gcomp.eps

\begin{figure}
   \epsfxsize = 3.0in
   \centerline{\epsfbox{fig10_bm.eps}}
   \caption{Various components of the energy, as computed in the
            $Z$-method, as a function of time. 
            Here we have introduced the notations:
            $E_a = $ 
            ${\displaystyle \frac{1}{2}} \dot A^2(t)$, 
            $E_b = $ 
            ${\displaystyle \frac{1}{2iN}}$
                   $\sum_{a=1}^{N}$
                   $
%                   {\displaystyle 
                    \left . 
%                           \partial^2 {\cal G}_{aa}(t,t') /
                          \frac{\partial^2 {\cal G}_{aa}(t,t') }
                               {\partial t \, \partial t'} \right |_{t=t'} 
%                   }
                   $, 
            $E_c = $ 
            ${\displaystyle \frac{1}{2iN}}$
                   $\sum_{a=1}^{N}$
                   $( m^2$ $+$ $e^2 A^2)$ ${\cal G}_{aa}(t,t)$, 
            $E_d = $ 
            ${\displaystyle \frac{1}{2iN}}$
                   $
%                   {\displaystyle 
                    \left . 
%                         \partial^2 D(t,t') /
                          \frac{\partial^2 D(t,t') }
                               {\partial t \, \partial t'} \right |_{t=t'} 
%                   }
                   $, 
            $E_e = $ 
            ${\displaystyle - \frac{1}{2iN^2}}$
                   $\sum_{a=1}^{N}$ $e^2$ $G_{aa}(t,t)$ $D(t,t)$,  
            $E_f = $ 
            ${\displaystyle - \frac{1}{N}}$ $e^2$ $A(t)$ $K^3(t,t; t)$.
           }
\label{fig:E_Gcomp}
\end{figure}

% Fig. 11: E_Zcomp.eps

\begin{figure}
   \epsfxsize = 3.0in
   \centerline{\epsfbox{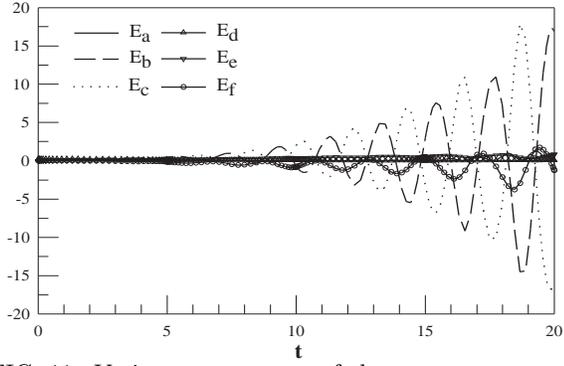}}
   \caption{Various components of the energy, as computed in the
            $Z$-method, as a function of time.
            The notations are similar to those of
            Fig.~\protect{\ref{fig:E_Gcomp}}, 
            with the addition that now we have
            ${\cal G}_{aa}(t,t') = $
            $G_{0 \, aa}(t,t')$ $+$ $G_{1 \, aa}(t,t') / N$, 
             and $G_{aa}(t,t')$ becomes $G_{0 \, aa}(t,t')$.
           }
\label{fig:E_Zcomp}
\end{figure}

%
%%%%%%%%%%%%%%%%%%%%%%%%%%%%%%%%%%%%%%%%%%%%%%%%%%%%%%%%%%%%%%
%

\end{document}

%% file: title2.tex
\catcode`\@=11

\def\maketitle2{\par % Uses \twocolumn[\@maketitle2].
\begingroup
\let\cite\@bylinecite
\def\thefootnote{\fnsymbol{footnote}}%
\twocolumn[\@maketitle2\vskip2pc]%
\thispagestyle{plain}\@thanks
\endgroup
\def\thefootnote{\arabic{footnote}}%
\setcounter{footnote}{0}%
\let\maketitle2\relax \let\@maketitle2\relax
\let\@thanks\relax \let\@authoraddress\relax \let\@title\relax
\let\@date\relax \let\thanks\relax \let\@abstract\relax 
\let\@pacs\relax}

\def\abstract#1{\gdef\@abstract{{\par % Store abstract text. 
\bgroup
\ifdim\prevdepth=-1000pt \prevdepth0pt\fi
\hsize\columnwidth
\dimen0=-\prevdepth \advance\dimen0 by17.5pt \nointerlineskip
\small\vrule width 0pt height\dimen0 \relax}{~~}#1\egroup}}

\def\pacs#1{\gdef\@pacs{{\par % Store PACS numbers as \@pacs.
\bgroup
\hsize\columnwidth \parindent0pt
\ifdim\prevdepth=-1000pt \prevdepth0pt\fi
\dimen0=-\prevdepth \advance\dimen0 by20pt\nointerlineskip
\egroup} PACS numbers:~#1}}

\def\@maketitle2{% Puts \@abstract and \@pacs in a {list}.
\@preprint
\@title
\ifdim\prevdepth=-1000pt \prevdepth0pt\fi
\@authoraddress
\@date
\begin{list}{}{\leftmargin=0.10753\textwidth \rightmargin=\leftmargin
\itemsep=1pc\partopsep=-1pc}
\item\@abstract
\item\@pacs
\end{list}
}

\catcode`\@=12